\documentclass[twoside]{article} 
\usepackage{fleqn,espcrc2} 
\usepackage{psfig}

\renewcommand{\thefootnote}{\fnsymbol{footnote}}

\title{Cumulant ratios in fully developed turbulence
\thanks{
Proceedings of the IX-th International Workshop on 
Multiparticle Production, Torino, June 11-18, 2000,
edited by  A.\ Giovannini and R.\ Ugoccioni;
Nuclear Physics B Supplement (to be published).}
}

\author{Hans C.\ Eggers\address[STL]{
                   Department of Physics, University of Stellenbosch,
                   7600 Stellenbosch, South Africa}
 and Martin Greiner\address[MPI]{
                   Max-Planck-Institut f\"ur Physik komplexer Systeme,
                   N\"othnitzer Str.\ 38, D--01187 Dresden, Germany}
                   \address[TUD]{
                Institut f\"ur Theoretische Physik, Technische Universit\"at,
                   D--01062 Dresden, Germany }
}

\begin{document}

\begin{abstract}
  In the context of random multiplicative cascade processes, we derive
  analytical solutions for one- and two-point cumulants with restored
  translational invariance. On taking ratios of cumulants in
  $\ln\varepsilon$, geometrical effects due to spatial averaging
  cancel out. These ratios can successfully distinguish between
  splitting functions while multifractal scaling exponents and
  multiplier distributions cannot.
\end{abstract}

\maketitle

\renewcommand{\thefootnote}{\arabic{footnote}}
\setcounter{footnote}{0}

\section{Introduction}

The Navier-Stokes equation governing fluid flow is deterministic.
Nevertheless, the statistical description of fully developed
turbulence has a long tradition \cite{MON71}. Random multiplicative
cascade models form a particularly simple and robust class of such
statistical models, reproducing important observed features such as
multiplier distributions and their correlations
\cite{SRE95,JOU99,JOU00} and related Kramers-Moyal coefficients
reflecting Markovian properties \cite{NAE97}. The models have worked
almost too well in the sense that different cascade-generating
probability density functions (pdf's or ``splitting functions'')
$p(q_L,q_R)$ for the multiplicative weights have been equally
successful in reproducing these observables. More sophisticated ways
to distinguish between them are clearly desirable.

While some experiments have concentrated on measuring statistics in
the energy dissipation density $\varepsilon$, we recently found a
complete analytical solution working in $\ln\varepsilon$ rather than
$\varepsilon$ itself \cite{GRE98,EGG98}.  We here and in Ref.\ 
\cite{EGG00a} show that cumulants in $\ln\varepsilon$ are analytically
calculable even when translational invariance is restored in order to
emulate the spatial homogeneity of experimental turbulence statistics.
Both one- and two-point cumulants turn out to be powerful tools which
for third and fourth order differ not only in magnitude but even in
sign for splitting functions which are indistinguishable in terms of
other observables.  Unlike multifractal scaling exponents, for
example, such cumulants can be expected to distinguish between
different models for sufficiently large experimental samples.

\section{Analytical solution for random multiplicative cascades}

Energy flux densities $\varepsilon$ are generated in the simplest
multiplicative cascade models as follows. In successive steps $j =
1,\ldots, J$, the integral scale $L$ is divided into equal intervals
of length $l_j = l_{j-1}/2 = L/2^j$ and dyadic addresses ${\bf \kappa}
= (k_1 \cdots k_j)$ with $k_i=0$ or $1$. At each step $j$, the energy
flux density $\varepsilon_{k_1 \cdots k_j}$ generates fluxes
multiplicatively in the two subintervals via
\begin{equation}
\label{nlya}
  \varepsilon_{k_1 \cdots k_j k_{j+1}}
     =  q_{k_1 \cdots k_j k_{j+1}} \;  \varepsilon_{k_1 \cdots k_j} 
     \,,
\end{equation}
where the random variables $q_L = q_{k_1 \cdots k_j 0}$ and $q_R =
q_{k_1 \cdots k_j 1}$ for the left and right subintervals are drawn
from a given cascade-generating pdf $p(q_L,q_R)$, independently of
other branches and generations of the dyadic tree.  When after $J$
cascade steps the smallest scale $\eta = l_J = L/2^J$ is reached, the
local amplitudes of the flux density field
\begin{equation}
\label{nlyb}
\varepsilon_t
= \varepsilon_{k_1 \cdots k_J}
= \prod_{j=1}^J   q_{k_1 \cdots k_j}
\end{equation}
at positions $0 < t(\kappa){=}(1+\sum_{j=1}^{J} k_j
2^{J-j}) \leq 2^J$ in units of $\eta$
are interpreted as the energy dissipation
amplitudes  which are to be compared to
experimental time series converted to one-dimensional spatial series
by Taylor's frozen flow hypothesis.

We have shown previously \cite{GRE98} that, since the product of
multiplicative weights (\ref{nlyb}) becomes additive on taking the
logarithm,
\begin{equation}
\ln\varepsilon_{k_1 k_2 \cdots k_J} = \sum_{j=1}^J \ln
q_{k_1 \cdots k_j} \,,
\end{equation}
the multivariate cumulant generating function for $\ln\varepsilon$ has
the analytical solution
\begin{eqnarray}
\label{nlyc}
  &&\!\!\!\!\!\!\!\!\!\!\!\!\!\!\!\!
   \ln Z(\lambda_{0 \cdots 0},\ldots,\lambda_{1 \cdots 1})
   \\
    &=&  \ln
         \bigg\langle  \exp\bigg(
         \sum_{k_1,\ldots,k_J=0}^1
         \lambda_{k_1 \cdots k_J}  \ln\varepsilon_{k_1 \cdots k_J}
         \bigg)  \bigg\rangle
         \nonumber \\
    &=&  \sum_{j=1}^J \; \sum_{k_1,\ldots,k_{j-1} = 0}^1 
         Q(\lambda_{k_1 \cdots k_{j-1}0} , \lambda_{k_1 \cdots k_{j-1}1})
         \nonumber \; ,
\end{eqnarray}
where the branching cumulant generating function $Q$ has arguments
\begin{equation}
\lambda_{k_1 \cdots k_j} 
 = \sum_{k_{j+1},\ldots,k_J = 0}^1  \lambda_{k_1 \cdots k_J} \,,
\end{equation}
(see Figure~1) and is defined by the Mellin transform of the
splitting function,
\begin{equation}
\label{nlyd}
  Q(\lambda_L,\lambda_R)
    =  \ln\left[ \int dq_L \, dq_R \, p(q_L,q_R) \, 
       q_L^{\lambda_L} \,  q_R^{\lambda_R} \right],
\end{equation}
Because of the simplicity of (\ref{nlyd}), $Q$ can often be found
analytically.

\begin{figure}[h]
\centerline{\psfig{file=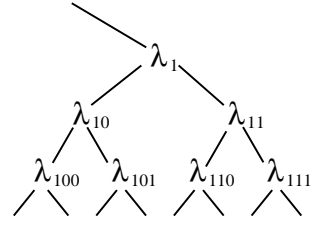,width=40mm}}
\caption{
The hierarchical structure of the cascade is reflected in
a corresponding structure in the source parameters $\lambda$.
}
\end{figure}

A host of analytical predictions for statistics in $\ln\varepsilon$
follow, starting with any and all $n$-multivariate cumulants obtained
directly from $\ln Z$ through
\begin{eqnarray}
\label{nlye}
  C(\kappa_1,\kappa_2,\cdots,\kappa_n)
&=& \langle 
     (\ln\varepsilon_{\kappa_1}) \cdots (\ln\varepsilon_{\kappa_n})
  \rangle_c
   \\
&=&  {\partial^n \ln Z  \over
        \partial\lambda_{\kappa_1} \partial\lambda_{\kappa_2}
        \cdots \partial\lambda_{\kappa_n}
       }\biggr|_{\lambda=0} 
     \nonumber
\end{eqnarray}
for arbitrary dyadic bin addresses $\kappa_1, \ldots , \kappa_n$.
These multivariate cumulants in $\ln\varepsilon$ are easily calculated
since, due to the additivity of $\ln Z$ in (\ref{nlyc}), they are
simple sums \cite{GRE98} of \textit{same-lineage cumulants} $c_n$ and
\textit{splitting cumulants} $c_{r,s}$ in $\ln q$ (see eqs.\ 
(\ref{rstf}) and (\ref{twpb}) below),
\begin{eqnarray}
\label{nlyf}
c_{n} 
    &=&  \left\langle (\ln q)^{n} \right\rangle_c
    \nonumber\\
    &=&  {\partial^{n} Q  \over  \partial \lambda_L^{n}
       }\biggr|_{\lambda_L=\lambda_R=0}
       \; , \\
\label{twoone8}
c_{r,s} 
    &=&  \left\langle (\ln q_L)^r(\ln q_R)^s \right\rangle_c
    \nonumber\\
    &=&  {\partial^{r+s} Q  \over  
        \partial \lambda_L^r  \partial \lambda_R^s
       }\biggr|_{\lambda_L=\lambda_R=0}
       \; ,
\end{eqnarray}
where without loss of generality we have assumed
$Q(\lambda_L,\lambda_R)$ to be symmetric in its arguments.

\begin{figure}[h]
\psfig{file=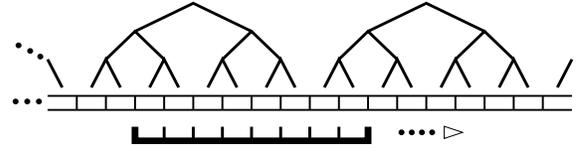,width=75mm}
\caption{
  Translational averaging using the moving window technique.
}
\end{figure}

\section{Restoring translational invariance}

Before the above theoretical cumulants can be compared to
experimentally measured ones, the issue of translational invariance
must be dealt with.  Clearly, the generating function (\ref{nlyc}) and
its cumulants (\ref{nlye}) are not translationally invariant, in
conflict with the homogeneous statistics characterising experimental
results.  Spatially homogeneous statistics can, however, be emulated
by creating a theoretical time series consisting of a chain of $m$
adjacent independent cascade fields with $2^J$ finest-scale bins each
\cite{GRE97}. In analogy to the experimental situation, an observation
window of width $2^J\eta$ is successively moved over this series in
bin-sized steps, $t=1,\ldots,(m-1)2^J$, successively ``seeing'' parts
of adjacent cascade configurations: see Figure 2.

A translationally
invariant one-point moment density would thus be constructed as
\begin{equation}
\label{rstb}
\overline{\rho}_n = \lim_{m\rightarrow\infty} \; {1\over (m-1)\,2^{J}}
\sum_{t=1}^{(m-1)2^J} \left( \ln\varepsilon_t \right)^n \,,
\end{equation}
which should be comparable to the experimental time series.
Operationally, this can be implemented by keeping only one cascade
while averaging over many different cascade configurations, i.e.\ 
\begin{equation}
\label{rstc}
\overline{\rho}_n 
= {1\over M} \sum_{t=1}^M 
  \left\langle ( \ln\varepsilon_t )^n \right\rangle 
= {1\over M} \sum_{t=1}^M
  \rho_n(t)
\,,
\end{equation}
with $M \equiv 2^J = L/\eta$ and $\langle \; \rangle$ denoting
configuration averaging. Likewise, a translationally invariant
two-point density with constant distance $\eta\,d$ (with $d =
1,2,3,\ldots$) between the two bins would be simulated by two adjacent
cascades,
\begin{equation}
\label{rstd}
\overline{\rho}_{r,s}(d) 
= {1\over M} \sum_{t=1}^M \rho_{r,s}(t,t{+}d) \,,
\end{equation}
with $\rho_{r,s}(t,t{+}d) = \left\langle (\ln\varepsilon_t)^r
  (\ln\varepsilon_{t+d})^s \right\rangle$. As shown in Figure~3, bin
$t{+}d$ at some stage in the summation exceeds $M = 2^J$ and hence
would refer to the right-hand cascade while $t$ would refer to the
left-hand one.
\begin{figure}[h]
\psfig{file=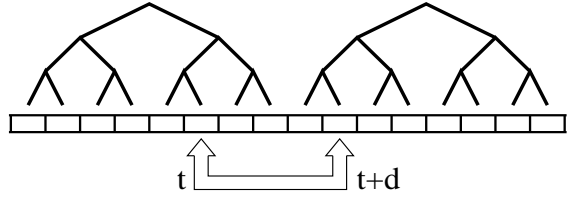,width=75mm}
\caption{
  Translational averaging for two-point statistics. Indices $t$ and
  $t{+}d$ run over all bins with the ``euclidean distance'' $d$ kept
  fixed.  }
\end{figure}

Given that the model provides solutions in terms of cumulants, it is
tempting to apply this averaging prescription directly to cumulants
also, i.e.\ to define
\begin{equation}
\label{rste}
\overline{C}_{r,s}(d) = {1\over M} \sum_{t=1}^M  C_{r,s}(t,t{+}d)
\,,
\end{equation}
using for $C_{r,s}(t,t{+}d)$ theoretical expressions obtained from
(\ref{nlye}). However, experimental cumulants are derived from
measured moments rather than the other way round \cite{CAR90} (for
example $C_2 \equiv \rho_2 - \rho_1^2$, $C_3 \equiv \rho_3 - 3\rho_2 +
2\rho_1^3$), so that averaging over moments rather than cumulants is
mandatory for theory also. The proper procedure is hence to convert
theoretical cumulants (\ref{nlye}) to moments, average these over $t$,
and then convert these back to cumulants for experimental comparison.

For the one-point case, this convoluted route becomes simple: the
$n$-th order one-point cumulant $C_n(t) \equiv C(\kappa_1 {=} \ldots
{=} \kappa_n{=}t)$, given by
\begin{equation}
\label{rstf}
  C_n(t) =  Jc_n \,,
\end{equation}
is independent of position $t$ so that translational averaging is
trivial. The only remaining complication is that $J$ is not an
experimental observable, and this is easily addressed by looking at
cumulant ratios $C_n/C_{n^\prime}$ for which the $J$-dependence
cancels. Ratios of translationally averaged one-point cumulants,
\begin{equation}
\label{rstg}
  { \overline C_n \over \overline C_{n^\prime} }
    =  { c_n \over c_{n^\prime} }
    =  { \langle \left( \ln q \right)^{n} \rangle_c  
         \over
         \langle \left( \ln q \right)^{n^\prime} \rangle_c }
       \;,
\end{equation}
being independent of $J$, should hence be directly comparable to
experiment.

To demonstrate the quality of these cumulant ratios, we consider three
model distributions, all with factorised splitting function
\begin{equation}
\label{rsth}
  p(q_L,q_R)
    =  p(q_L) \, p(q_R) \,,
\end{equation}
namely a binomial distribution (often also termed the
``$\alpha$ model''),
\begin{eqnarray}
\label{rstj}
  p(q)
    &=&  {\alpha_2 \over {\alpha_1+\alpha_2}} \;
       \delta\left( q-(1-\alpha_1) \right)
       \nonumber\\
    &+& {\alpha_1 \over {\alpha_1+\alpha_2}} \;
       \delta\left( q-(1+\alpha_2) \right)
       \,,
\end{eqnarray}
with parameters $\alpha_1=0.3$ and $\alpha_2=0.65$, a log-normal
distribution
\begin{equation}
\label{rstk}
  p(q)
    =  {1\over \sqrt{2\pi} \; \sigma q}
       \exp\left[ - {1 \over 2 \sigma^2}
         \left( \ln q + \frac{\sigma^2}{2} \right)^2 
       \right]
\end{equation}
with parameter $\sigma=0.42$, and a beta distribution
\begin{equation}
\label{rstl}
  p(q)
    =  { \Gamma(\beta_1+\beta_2) \over
         \Gamma(\beta_1)\Gamma(\beta_2) } \,
       8^{1-\beta_1-\beta_2} \,
       q^{\beta_1-1} \, (8-q)^{\beta_2-1}
\end{equation}
with parameters $\beta_1=4.88=\beta_2/7$ and $q \in (0,8)$. The beta
model is particularly appealing because it parallels the experimental
situation where energy conservation in three dimensions results in a
non-energy-conserving one-dimensional projection. Parameter values
quoted are the result of requiring $\langle q \rangle = 1$ and best
fits needed to reproduce observed multiplier statistics \cite{SRE95},
which hence cannot distinguish between these three models.

\begin{figure}[h]
\psfig{file=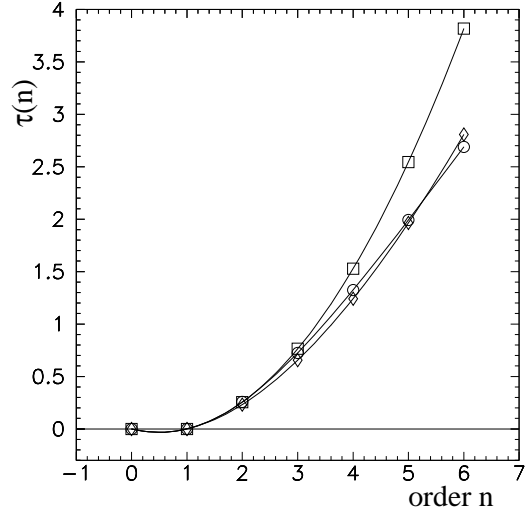,width=75mm}
\caption{
Multifractal scaling exponents $\tau(n)$ for the
binomial (circles, eq.\ (\ref{rstj})), 
lognormal (squares, eq.\ (\ref{rstk})) 
and beta (diamonds, eq.\ (\ref{rstl})) distributions.
}
\end{figure}

\begin{figure}[h]
\psfig{file=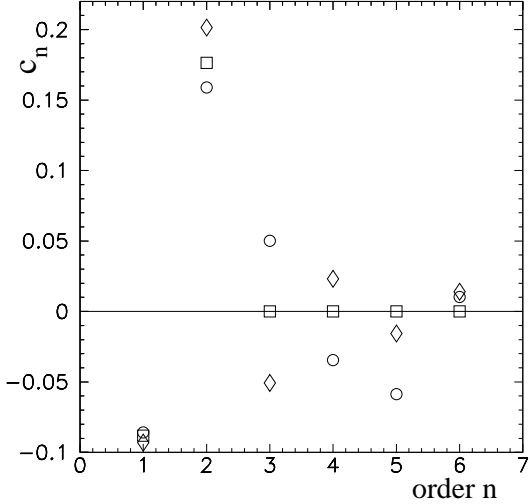,width=75mm}
\caption{
Same-lineage cumulants $c_n$ for the
binomial (circles, eq.\ (\ref{rstj})), 
lognormal (squares, eq.\ (\ref{rstk})) 
and beta (diamonds, eq.\ (\ref{rstl})) distributions.
}
\end{figure}

As shown in Figure~4, all three splitting functions also have almost
identical multifractal scaling exponents $\tau(n) = \ln\langle
q^n\rangle / \ln 2$. Since $\langle q \rangle = 1$ by construction,
$\tau(1)$ is zero for all three distributions.  For $n=2$ we get
$\tau(2) = 0.26$ for the first two distributions and $0.23$ for the
beta distribution, indistinguishable within the uncertainty of the
experimental intermittency exponent $\mu = 0.25 \pm 0.05$
\cite{SRE97}. We secondly note that even for $n \geq 3$ the
$\tau(n)$'s for the binomial and the beta distributions remain
indistinguishable.  Thirdly, all three distributions have a positive
skewness $\langle (q-\langle q\rangle)^3 \rangle / \langle (q-\langle
q\rangle)^2 \rangle^{3/2}$ when measured in $q$ and reproduce the
observed multiplier statistics, including their correlations.  This
has been shown for the binomial and log-normal in Ref.~\cite{JOU00}.
Numerical analysis of the beta distribution yields similar results.
Fits to observed scaling exponents have not been performed because it
is not straightforward to compare theoretical and experimental scaling
exponents due to the finiteness of the inertial range \cite{SCH99}.

The above observables thus fail manifestly to distinguish between the
different model distributions. By contrast, Figure~5 demonstrates that,
while $c_1$ and $c_2$ are almost identical for the three models,
higher-order cumulants and cumulant ratios (\ref{rstg}) are very
different.  For example, $c_3=0.05$, $0.00$ and $-0.05$ for the
distributions (\ref{rstj}), (\ref{rstk}) and (\ref{rstl})
respectively, so that the theoretical cumulant ratios
\begin{eqnarray}
\label{rstm}
  {\overline{C}_3 \over \overline{C}_2 }
    &=&{c_3 \over c_2}
    \; = \;  {\left\langle (\ln q)^3 \right\rangle_c \over 
        \left\langle (\ln q)^2 \right\rangle_c }
      \\
    &=&  \left\{
       \begin{array}{rl}
           0.31  &   \qquad {\rm (binomial)}   \\
           0.00  &   \qquad {\rm (lognormal)}  \\
          -0.25  &   \qquad {\rm (beta)}       
       \end{array}
       \right.
       \nonumber
\end{eqnarray}
lead to results that differ even in sign. If the present model
assumptions are adequate, this sign difference should be seen in the
experimental ratio
\begin{equation}
\label{rstn}
  {\overline C_3^{\rm obs} \over \overline C_2^{\rm obs}}
=
  {   \langle (\ln\varepsilon)^3 \rangle
   - 3\langle (\ln\varepsilon)^2 \rangle
      \langle  \ln\varepsilon    \rangle
   + 2\langle  \ln\varepsilon    \rangle^3
  \over
      \langle (\ln\varepsilon)^2 \rangle
    - \langle  \ln\varepsilon    \rangle^2
  } \,.
\end{equation}
In fourth order, the same-lineage cumulant $c_4$ has a different
sign for the binomial and beta distribution, so that the ratios
$\overline{C}_4 / \overline{C}_2 = c_4 / c_2$ of two-point cumulants
come with a different sign, too. For the log-normal distribution,
these ratios are of course again zero.

\begin{figure}[h]
\centerline{\psfig{file=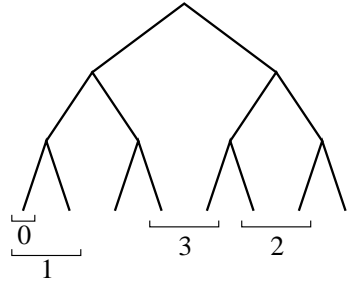,width=45mm}}
\caption{
Some examples for the ultrametric distance $D$ in a cascade
with $J=3$.
}
\end{figure}

\section{Two-point cumulants and geometry}

Having demonstrated the advantages of measuring ratios of one-point
cumulants, we now consider equivalent two-point ratios. The
theoretical two-point cumulant for two bins is found in terms of their
mutual ultrametric distance $D > 0$. As illustrated \cite{EGG98} in
Figure~6, two bins $\kappa_1=(k_1 \cdots k_j k_{j+1} \cdots k_J)$ and
$\kappa_2=(k_1 \cdots k_j k_{j+1}^\prime \cdots k_J^\prime)$, with
$k_{j+1} \neq k_{j+1}^\prime$, are separated by an ultrametric
distance $D=J-j$.
\begin{equation}
\label{twpb}
C_{r,s}(D) = (J-D) c_{r+s} + c_{r,s} \;.
\end{equation}
Again we must restore translational invariance using (\ref{rstd}) for
theoretical densities $\rho_{r,s}$ rather than (\ref{rste}) for
cumulants. In doing so, we must note that, since the density
$\rho_{r,s}(t,t{+}d)$ factorises when bins $t$ and $t+d$ belong to
independent cascades, i.e.\ whenever $M-d+1 \leq t \leq M$, the
averaged moment splits up,
\begin{eqnarray}
\label{twpc}
\overline \rho_{r,s}(d)
&=& {1\over M} \sum_{t=1}^{M-d} \rho_{r,s}(t,t+d)
\nonumber \\
&+& {1\over M} \sum_{t=M-d+1}^M \rho_{r}(t) \rho_{s}(t+d) \,.
\end{eqnarray}
Analytic expressions for $\rho_{r,s}(t,t+d)$ are again readily derived
by inserting the cumulants (\ref{rstf}) and (\ref{twpb}) into
the usual relations between $n$-variate moments and cumulants
\cite{CAR90} and thence into (\ref{twpc}).

Since the cumulants $c_n$ and $c_{r,s}$ are independent of $t$, this
procedure clearly involves summation of $(J-D)$ over $t$ as in
(\ref{twpb}) to create ``geometrical coefficients'' of $c_n$ and
$c_{r,s}$ of the type
\begin{equation}
\label{twpd}
  G_n(J,d) 
    =  {1\over M}
       \sum_{t=1}^{M - d} \left(J - D(t,t+d)\right)^n
\end{equation}
with $n = 0, 1, 2, \ldots$, where the dependence of the ultrametric
distance $D$ on the bin positions is made explicit. These coefficients
are best evaluated by changing the index of of summation,
\begin{equation}
\label{twpf}
  G_n(J,d) 
    =  \sum_{D=1}^J p(D|J,d) \, (J-D)^n  \; ,
\end{equation}
with $p(D|J,d)$ the (normalised) histogram function counting the
number of times $D$ appears while $t$ runs over its allowed values.
Empirically, we find
\begin{equation}
\label{twpg}
  p(D|J,d) 
    =  \left\{
       \begin{array}{lrl}
         0                  &   (1 \leq D < A) & \\
         1 - (d/2^A)        & (D = A)          & \\
         d/2^D              &   (A <  D \leq J) & ,
       \end{array}
       \right.
\end{equation}
where $A = \lceil \log_2 d \rceil$ is the ceiling of $\log_2 d$.
Insertion of (\ref{twpg}) into (\ref{twpf}) leads to analytical
expressions for the geometrical coefficients
\begin{eqnarray}
\label{twph}
  G_0(J,d) 
    &=&  (1 - 2^{-J}d)
         \; ,  \\
\label{twphb}
  G_1(J,d) 
    &=&  (J-A) - 2d(2^{-A} - 2^{-J})
         \; ,  \\
\label{twphc}
  G_2(J,d) 
    &=&  (J-A)^2 - 4d(J-A)2^{-A} \nonumber\\
    &+&  6d(2^{-A} - 2^{-J}) \,,
\end{eqnarray}
which in turn yield analytical results for the averaged two-point
densities $\overline{\rho}_{r,s}(d)$ of (\ref{twpc}).  Spatially
homogeneous two-point cumulants are then constructed via the inversion
formulae \cite{CAR90}
\begin{eqnarray}
\label{twpja}
  \overline{C}_{1,1}(d)
    &=&  
           \overline{\rho}_{1,1}(d)
         - \overline{\rho}_{1}^2
         \,,
         \\
\label{twpjb}
  \overline{C}_{2,1}(d)
    &=&  
             \overline{\rho}_{2,1}(d)
         - 2 \overline{\rho}_{1} \overline{\rho}_{1,1}(d)
              \nonumber\\ &&
         -\; \overline{\rho}_{2} \overline{\rho}_{1}
         + 2 \overline{\rho}_{1}^3
         \,,
         \\
\label{twpjc}
  \overline{C}_{3,1}(d)
    &=&  
             \overline{\rho}_{3,1}(d)
         - 3 \overline{\rho}_{1} \overline{\rho}_{2,1}(d) 
         -   \overline{\rho}_{3} \overline{\rho}_{1}
              \nonumber\\ &&
         -\; 3 \overline{\rho}_{2} \overline{\rho}_{1,1}(d)
         + 6 \overline{\rho}_{1}^2 \overline{\rho}_{1,1}(d)
              \nonumber\\ &&
         +\; 6 \overline{\rho}_{2} \overline{\rho}_{1}^2
         -   6 \overline{\rho}_{1}^4
         \,,
         \\
\label{twpjd}
  \overline{C}_{2,2}(d)
    &=&
             \overline{\rho}_{2,2}(d)
         - 4 \overline{\rho}_{1} \overline{\rho}_{2,1}(d)
         - 2 (\overline{\rho}_{1,1}(d))^2
              \nonumber\\ &&
         -\;   \overline{\rho}_{2}^2
         + 8 \overline{\rho}_{1}^2 \overline{\rho}_{1,1}(d)
              \nonumber\\ &&
         +\; 4 \overline{\rho}_{2} \overline{\rho}_{1}^2
         - 6 \overline{\rho}_{1}^4 
         \,.
\end{eqnarray}
With Eqs.\ (\ref{twpc})--(\ref{twphc}), we arrive for $s{=}1$ at
\begin{equation}
\label{twpk}
  \overline C_{r,1}(d)
    =  G_1(J,d) \, c_{r{+}1} + G_0(J,d) \, c_{r,1} \,.
\end{equation}
This turns out to be equivalent to direct translational averaging of
cumulants (\ref{rste}). For $s{\neq}1$, however, such direct averaging
is wrong and the full conversion from cumulant to moment to averaged
moment and back to averaged cumulant is unavoidable. For $r{=}s{=}2$
we get, for example,
\begin{eqnarray}
\label{twpl}
\overline C_{2,2}(d) 
&=&  G_2(J,d) c_{2}^2
   +  G_1(J,d) \left( c_{4} + 4c_{2}c_{1,1}  \right)
       \nonumber \\
&+&  G_0(J,d) \left( c_{2,2} + 2c_{1,1}^2 \right) 
       \nonumber \\
&-& 2 \left[ G_1(J,d) c_{2} + G_0(J,d) c_{1,1} \right]^2
       \,, 
\end{eqnarray}
where the additional terms are a consequence of the third term in the
expression (\ref{twpjd}) for $\overline C_{2,2}(d)$. Averaged
two-point cumulants $\overline C_{r,s}$ of higher order $r,s{\geq}2$
exhibit similar structures. Translationally averaged $n$-point
cumulants $\overline C_{m_1,\ldots,m_n}$ can be calculated by the
procedure sketched above.

Figure 7 shows explicit examples for $G_0$ and $G_1$ for a cascade of
length $J=5$. As expected, $G_0$ reflects the trivial dependence of
the splitting cumulant on the sum limits $1 \leq t \leq M{-}d$, apart
from the point $d{=}0$ for which no splitting cumulant enters at all.
More interesting is the coefficient for the same-lineage cumulant,
$G_1$: it consists of a series of straight-line segments, changing
slope whenever $d\,\textrm{mod}\, 2 = 0$ and ending at zero for $d
\geq 2^J/2$.  Since the form of $G_1$ changes whenever $d$ is a power
of 2, approximate exponential behaviour of $\overline{C}_{r,1}$ as a
function of $d$ is to be expected; this is shown in Figure 8. The
exponential form would, however, be destroyed by any sizeable
contribution of $G_0$ entering via the splitting cumulant, especially
at larger $d$.

The form of $G_1$ can be further understood by considering an
alternative formulation for $p(D|J,d)$ in terms of $k = \lfloor \log_2
d \rfloor$,
\begin{eqnarray}
\label{twpm}
p(D|J,d) 
&=& (1 - d 2^{-J}) \delta_{J-D} \\
&+& \sum_{j=1}^{J-1} \Theta(J-j-k)\; (1 - 2^{j-J} \,d)
   \nonumber\\
&& \qquad \times  ( \delta_{J-D-j} - \delta_{J-D-j+1} ) \,, \nonumber
\end{eqnarray}
(with $\delta_n = \delta_{n,0}$ the Kronecker delta and 
$\theta(n) = 0$ whenever $n\leq 0$ and 1 otherwise)
since, with $G_1 = \sum_{D=1}^{J} (J-D) \, p(D|J,d)$, we find
\begin{equation}
\label{twpn}
G_1(J,d) = \sum_{j=1}^{J-1} \Theta(2^{J-j} - d) \; (1 - 2^{j-J} \,d) 
\,,
\end{equation}
which is a sum of straight-line contributions kicking in whenever $d$
becomes smaller than $2^{J-j}, j = 1,2,\ldots$ This means that
whenever $d$ becomes smaller than some dyadic fraction of $M$, the two
bins $t$ and $t{+}d$ can fall within the same $(J{-}j)$-scale
subcascade so that $G_1$ picks up new contributions from this scale.

\begin{figure}
\psfig{file=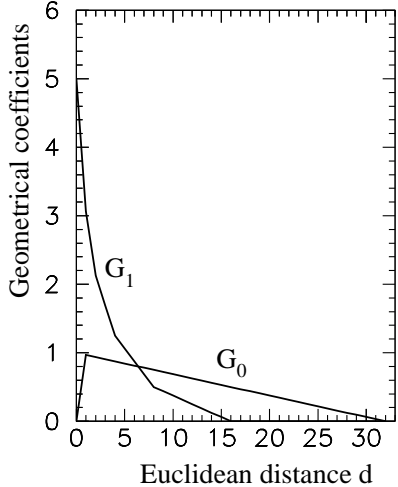,width=66mm}
\caption{
  Geometrical coefficients $G_1$ and $G_0$, on linear scale for
  $J{=}5$.  }
\end{figure}

Translationally invariant cumulants $\overline{C}_{r,s}$ are
constructed from these factors according to eqs.\
(\ref{twpk})--(\ref{twpl}). Figure~9 shows by example
$\overline{C}_{1,1}$ for the binomial ($\alpha$ model) and the
corresponding energy-conserving $p$-model with pdf
\begin{eqnarray*}
p(q_L,q_R) 
&=& \left[\textstyle{1\over2} \delta(q_L - 1 - \alpha)
        + \textstyle{1\over2} \delta(q_L - 1 + \alpha) \right]
      \nonumber\\
&&\; \times\; \delta(q_L + q_R - 2) \,,
\end{eqnarray*}
setting (for purposes of comparison) $\alpha_1 = \alpha_2 = \alpha
= 0.4$. The $\alpha$-model has $c_{r,s} = 0$ and hence contains no
contribution from $G_0$ but only from $G_1$ and $c_2 =
\textstyle{1\over 4} \left[ \ln\left((1+\alpha) / (1 - \alpha)\right)
\right]^2 = 0.1795$, while for the $p$-model both $c_{1,1} = {-}c_2$
and $c_2$ contribute in (\ref{twpk}).  The resulting $p$-model
$\overline{C}_{1,1}$ has the same peak at $d{=}0$ as the
$\alpha$-model but exhibits the familiar anticorrelation (negative
cumulant) at larger $d$ \cite{GRE96}. Whether and for what $d$ the
$\overline{C}_{1,1}$ is negative depends, however, on the sum of
same-side and splitting cumulant contributions rather than on the
splitting cumulant alone.

\begin{figure}
\hspace*{-5mm}\psfig{file=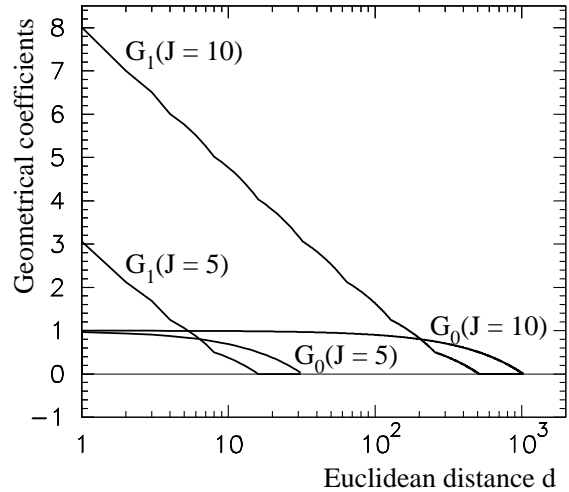,width=88mm}
\caption{
  Geometrical coefficients $G_1$ and $G_0$, on logarithmic scale for
  $J{=}5$ and $J{=}10$, showing the approximately exponential
  behaviour of $G_1$.}
\end{figure}

\begin{figure}[h]
\begin{center}
\parbox[t]{100mm}{\psfig{file=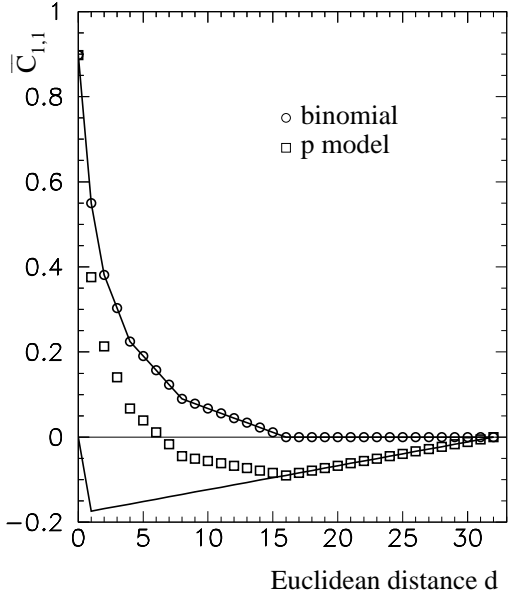,width=78mm}}
\end{center}
\caption{
  $C_{1,1}$ two-point cumulants for the binomial and $p$-models with
  $J=5$, showing contributions from same-lineage and splitting parts
  of the respective splitting functions. Solid lines represent the
  (scaled) geometrical coefficients.  }
\end{figure}

We further note that \textit{all} models whose splitting function
factorises have zero translationally invariant two-point cumulants for
$d\geq 2^J/2$. Roughly, this can be translated into the statement that
deviations of two-point cumulants from zero for ``long'' distances $d$
(compared to an admittedly fluctuating cascade size which we have
modelled as a constant $2^J$) would signal the nonfactorisation of the
splitting function and vice versa.

Returning to cumulant ratios, we focus on cumulants $\overline
C_{r,1}(d)$. If the splitting function factorises as in (\ref{rsth}),
then the splitting cumulant $c_{r,s}$ is zero and the two-point
cumulant for $d\geq 1$, becomes directly proportional to the
geometrical coefficient $G_1(J,d)$,
\begin{equation}
\label{twpp}
  \overline C_{r,1}(d) 
    =  c_{r{+}1} \, G_1(J,d)  \,.
\end{equation}
Taking ratios of two-point cumulants of different orders,
\begin{equation}
\label{twpq}
  { \overline C_{r,1}(d) \over \overline C_{r^\prime,1}(d) }
    =  { c_{r{+}1} \over c_{r^\prime{+}1} }
    =  { \langle ( \ln q )^{r+1} \rangle_c  
         \over
         \langle ( \ln q )^{r^\prime+1} \rangle_c }
       \; ,
\end{equation}
the geometrical coefficient drops out, so that these ratios become
independent of $d$. This is an important observation as it grants
access to properties of the pdf (cascade generator) even after spatial
homogeneity has been restored.  Also, the $d$-independence of these
ratios constitutes a severe test of the model assumptions entering the
cascade models.  Furthermore, the factorisation assumption can be
tested since one- and two-point ratios are equal if (\ref{rsth})
holds,
\begin{equation}
\label{twpr}
{ \overline C_{r,1}(d) \over \overline C_{r^\prime,1}(d) }
=  { c_{r{+}1} \over c_{r^\prime{+}1} }
=  { \overline C_{r+1} \over \overline C_{r^\prime+1} }
     \quad \forall\; d  \; .
\end{equation}
In this case, ${ \overline C_{2,1} / \overline C_{1,1}}$ would assume
the same numerical values as those in Eq.\ (\ref{rstm}), with similar
powers to discriminate between models. Two-point cumulant ratios would
hence also predict clearly different results, in contrast with scaling
exponents and multiplier distributions.

We also note that the connection \cite{GRE98} between the multifractal
scaling exponents $\tau(n{=}\lambda)$ and the cumulant branching
generating function (\ref{nlyd}),
\begin{equation}
\label{twps}
  \tau(\lambda) 
    = {\ln\left\langle q^\lambda \right\rangle \over \ln 2}
    = {Q(\lambda,0) \over \ln 2}
      \; ,
\end{equation}
implies that the same-lineage cumulants (\ref{nlyf}), which are to
be extracted from ratios (\ref{rstg}) or (\ref{twpp}), are related
to the $\tau(n)$ by
\begin{equation}
\label{twpt}
  c_n
    =  \left.
       {\partial^n Q(\lambda,0) \over \partial \lambda^n}
       \right|_{\lambda=0}
    =  \left. \ln 2 \;
       {\partial^n \tau(\lambda) \over \partial \lambda^n}
       \right|_{\lambda=0}
       \; ,
\end{equation}
i.e.\ the cumulant $c_n$ in $\ln q$ is related to the $n$-th
derivative of the scaling exponent $\tau(\lambda)$, taken at
$\lambda=0$. In principle, this not only allows for an unambiguous,
albeit indirect extraction of scaling exponents, but also of the more
fundamental splitting function.

We end on an interesting sideline regarding the detection of scaling
in the bin size $\ell$. Conventionally, this is done by plotting
$\ln\langle \varepsilon^n \rangle$ against $\ln \ell$ in the
expectation of seeing a straight line. The same scaling of $\langle
\varepsilon^n \rangle$ is just as easily detected by pointing out that
the one-point cumulant in $\ln\varepsilon$ is given at every scale $j$
by $\overline{C}_n^{(j)} = j c_n$ and, since $\ell_j/L = 2^{-j}$,
\begin{equation}
\label{twpv}
\overline{C}_n^{(j)} = {\ln(L/\ell_j)\over \ln 2}\; c_n
      \;;
\end{equation}
in other words, scaling in $\langle\varepsilon^n\rangle$ is manifest
in a logarithmic dependence on the length scale $\ell_j$ of the
corresponding one-point cumulant in $\ln\varepsilon$. It must
be remembered, though, that such ``forward'' scaling behaviour
can be destroyed by the processes of translational averaging as
well as the experimental ``backward'' box summation \cite{GRE96}.

\section{Discussion}

We have shown that features of the analytical solution for cumulants
in $\ln\varepsilon$ can be preserved beyond the complication of
translational invariance, and in the process elucidated the interplay
between the same-lineage and splitting cumulants generated at each
cascade splitting on the one hand, and the geometrical features on the
other. We are, of course, tempted to apply two-point cumulants of
$\ln\varepsilon$ directly to the experimental energy dissipation field
deduced from hot-wire time series and to study possible dependences on
the Reynolds number and the flow configuration. This may be done in
the spirit of naive discovery. We do think, however, that studies of
different model assumptions such as continuous multiplicative cascade
processes \cite{SCH97}, hierarchical shell models \cite{BEN97},
effects of finite inertial range etc.\ should sensibly be undertaken
before taking the comparison with data too seriously.
\\

\noindent\textbf{Acknowledgements:}\\
We thank J\"urgen Schmiegel for fruitful discussions.  This work was
funded in part by the South African National Research Foundation. HCE
thanks the organisers of this workshop and the MPIPKS for kind
hospitality and support.


\end{document}